%% file: main.tex
\documentclass{article}
\usepackage[T1]{fontenc}
\usepackage[preprint]{ismir} %
\usepackage{amsmath,cite,url}
\usepackage{graphicx}
\usepackage{tabularx}
\usepackage{booktabs}

\usepackage{xcolor}
\usepackage{booktabs}

\usepackage{todonotes}
\usepackage{multirow}
\usepackage{enumitem} 

\usepackage{svg}

\newcommand{\musicabench}[0]{MusICA-MetaBench}
\newcommand{\metaq}[0]{question template}

\newcommand{\noinput}[0]{\texttt{no-input}}
\newcommand{\noiseinput}[0]{\texttt{noise-input}}

\newcommand{\qsPerSubcategory}[1]{\fpeval{3 * 5 * #1}}

\definecolor{lightblue}{RGB}{173,216,230}
\definecolor{lightgreen}{RGB}{200,230,200}
\definecolor{lightyellow}{RGB}{255,250,205}
\definecolor{lightred}{RGB}{250,200,200}
\definecolor{lightpurple}{RGB}{230,210,250}
\definecolor{lightgray}{RGB}{240,240,240}
\definecolor{lightcyan}{RGB}{210,245,245}
\definecolor{lightmint}{RGB}{220,255,240}
\definecolor{lightlime}{RGB}{230,255,200}
\definecolor{lightorange}{RGB}{255,230,200}
\definecolor{lightpeach}{RGB}{255,218,185}
\definecolor{lightpink}{RGB}{255,220,230}
\definecolor{lightrose}{RGB}{255,200,220}
\definecolor{lightlavender}{RGB}{240,230,255}
\definecolor{lightindigo}{RGB}{220,220,255}
\definecolor{lightteal}{RGB}{200,240,235}
\definecolor{lightbrown}{RGB}{235,220,200}

\let\citet\cite
\let\citep\cite

\DeclareRobustCommand{\paragraph}[1]{%
  \textbf{#1}%
}

\newcommand{\toFinalSubmission}[1]{}

\title{Music I Care About: Automated Multimodal Benchmarking of LLM Music Perception Skills on (Almost) Any Music}
\threeauthors
  {Tomáš Sourada} {Charles University \\ \texttt{sourada@ufal.mff.cuni.cz}}
  {Katia Vendrame} {Brno University of Technology \\ \texttt{ivendrame@fit.vut.cz}}
  {Jan Hajič jr.} {Charles University \\ \texttt{hajicj@ufal.mff.cuni.cz}}

\def\authorname{T. Sourada, K. Vendrame, and J. Hajič jr.}

\usepackage[bookmarks=false,pdfauthor={\authorname},pdfsubject={\pdfsubject},hidelinks]{hyperref}
\usepackage{cleveref}

\usepackage[absolute]{textpos}

\begin{document}

\maketitle

\begin{abstract}
Music represents a cornerstone of human culture, existing digitally across diverse modalities, including audio, symbolic encodings (e.g., MIDI, MusicXML), and sheet music. Despite the advancement of Multimodal Large Language Models (MLLMs), current music benchmarks face three major limitations. 
First, large static benchmarks are resource-intensive to evaluate, and it remains unclear how their results transfer to diverse kinds of music beyond those included in the benchmark. Second, benchmarks claiming to measure ``music understanding'' often fail to require music perception. Third, they do not support systematic performance comparisons across musical modalities. To overcome these issues, we introduce the Music I Care About Meta-Benchmark (\textbf{MusICA-MetaBench}), a framework that automatically derives on-demand benchmarks directly from user-provided data.
By leveraging structured symbolic representations (e.g., MusicXML) and our pre-defined question templates, we build multiple-choice question-answer pairs that probe music perception competencies, aligned with music pedagogy, across audio, music notation images, and symbolic files. 
We demonstrate our framework with the ChoraleBricks dataset, and experimentally determine benchmark sizes that ensure statistically reliable model comparisons for this setup. By comparing against text-only and white-noise baselines, we show our questions do measure music perception. 
Ultimately, MusICA-MetaBench represents a significant advancement in 
the cross-modal assessment of music perception for MLLMs. By proposing a 
dataset-specific benchmarking paradigm, it enables efficient on-demand evaluation of music perception capabilities.
\end{abstract}

\section{Introduction}\label{sec:introduction}

Music exists digitally across three modalities: audio recordings, symbolic
encodings (e.g., MIDI, MusicXML, ABC notation), and sheet music images.
Multimodal Large Language Models (MLLMs) have demonstrated strong capabilities
in audio and visual understanding~\cite{openai2024gpt4ocard,comanici2025gemini25pushingfrontier,xu_qwen3-omni_2025},
and their application to music is an active area of MIR
research~\cite{DBLP:conf/ismir/MaLYBM25,zang_are_2025}. Evaluating how well these
models handle musical tasks is both timely and consequential: it determines
how much of our field's research agenda can be delegated to general-purpose
models, and where purpose-built MIR systems remain indispensable. Yet rigorous
evaluation across diverse repertoires, modalities, and research contexts
remains an open methodological challenge.

Existing benchmarks for evaluating MLLMs on music suffer from three major
limitations. First, large static benchmarks are impractical: they are
resource-intensive to evaluate; accuracy values age quickly as new models are
released; results may not transfer beyond the styles and genres included;
benchmark data may leak into future training corpora, invalidating comparisons;
human annotation of ground truth is extremely costly while LLM-assisted annotation raises
quality concerns; no single benchmark can cover music's diversity, and new
models arrive faster than evaluation can keep pace; and copyrighted material
creates legal constraints on what data can be included.

Second, some benchmarks claiming to evaluate music \textit{understanding}
(e.g., by asking questions about audio) do not require actual perception of
musical content, as their questions are answerable from textual cues
alone~\cite{zang_are_2025,yue_mmmu-pro_2025}.

Third, most benchmarks cover a single modality ---audio~\cite{weck_muchomusic_2024,zang_are_2025,weck_hummusqa_2026},
sheet music images~\cite{mundada_wildscore_2025,chen_musixqa_2025}, or
symbolic encodings~\cite{zhao_abc-eval_2025,yuan_chatmusician_2024} --- with
only a few recent exceptions spanning two modalities~\cite{wang_towards_2025,dai_musical_2025,zhao_museagent-1_2026}.
This prevents cross-modal comparison and leaves a key question
open: does poor MLLM performance on music reflect modality-specific perceptual
limitations, or deeper conceptual deficits independent of representation?

To address these limitations, we introduce the \textit{Music I Care About
Meta-Benchmark} (\musicabench{}). Rather than a fixed benchmark,
\musicabench{} automatically generates on-demand benchmarks from
user-provided musical data. Given a piece in symbolic representation
(MusicXML), we programmatically extract ground-truth answers for predefined
piece-level question templates using the \texttt{music21}
library~\citep{DBLP:conf/ismir/CuthbertA10}, yielding question--answer pairs grounded in
that specific piece.

\subsection{MusICA-MetaBench Design Principles}

Question templates are grounded in music-theory feature-recognition skills
(see \Cref{tab:meta-question-table}), framing music perception as the
identification and relation of pitch and rhythmic patterns, cross-culturally universal features~\cite{doi:10.1073/pnas.1414495112,mcbride_convergent_2023,jacoby_integer_2017}.

When input data is multimodal and piece-aligned (the same piece represented
simultaneously as audio, sheet music image, and symbolic file), the same
question can be posed across all three modalities with an identical
ground-truth answer, enabling direct cross-modal comparison on the same
underlying musical content.

We experimentally determine optimal benchmark sizes to be large enough to yield
statistically significant comparisons at a desired effect size, yet small
enough to keep evaluation fast and inexpensive. By comparing model
performance against a \noinput{} condition (question only, no musical file)
and a \noiseinput{} condition (random noise in place of the musical file),
we verify that solving the benchmark requires musical perception.

We demonstrate \musicabench{} by applying it to two multimodal datasets,
ChoraleBricks~\cite{balke_choralebricks_2025} and
ChoralSynth~\cite{narang_choralsynth_2023}, and evaluating 8 state-of-the-art
MLLMs on the generated benchmarks.

The intended use of \musicabench{} is not objective MLLM comparison, but
actionable diagnostics for stakeholders applying MLLMs to their own musical
data, for example, MIR researchers assessing model readiness for specific
tasks. By enabling evaluation on user-provided repertoire, \musicabench{}
reveals not only whether MLLMs fail, but where: across modalities, perceptual
dimensions, and repertoires.

The pipeline (benchmark generation and response
evaluation) is programmatic (LLM-independent), ensuring
controllability and interpretability, and restricting the role of (M)LLMs
solely to that of evaluation subjects.

We regard this as a first step rather than a complete solution, but believe
this paradigm makes MLLM evaluation on music faster, cheaper, domain-flexible,
and more interpretable. We release all code and generated benchmarks.\footnote{\url{https://github.com/tomsouri/MusICA-MetaBench-preprint}}

The paper is organized as follows:
\Cref{sec:related_work} surveys recent music benchmarks for LLMs;
\Cref{sec:methodology} describes the question templates and benchmark
generation pipeline;
\Cref{sec:exps-10times-methodology-calibration-of-benchmark-size} details
benchmark-size calibration;
\Cref{sec:normal-no_input-noise_input} demonstrates that perception is
required to solve the benchmark;
\Cref{sec:test-drive} demonstrates practical usage and provides guidelines
for custom datasets; and
\Cref{sec:discussion} concludes with contributions, limitations, and future
directions.

\section{Related Work}\label{sec:related_work}

Recent benchmarks for music understanding (2024–2025) predominantly evaluate (M)LLMs via question answering (QA)~\cite{weck_muchomusic_2024,zang_are_2025,koh_jamendo-qa_2025,dai_musical_2025,mundada_wildscore_2025,chen_musixqa_2025,wang_towards_2025,zhao_abc-eval_2025}, with multiple-choice (closed QA) being the dominant format.

A known weakness of closed QA is that some benchmarks can be partially solved without perceiving any musical content, through reasoning over answer options alone~\cite{zang_are_2025,yue_mmmu-pro_2025}. \citet{zang_are_2025} demonstrated that a text-only LLM reaches 56\% accuracy on MuChoMusic, far above the 25\% random baseline. 
Proposed mitigations include filtering out questions solvable by text-only LLMs~\cite{yue_mmmu-pro_2025} and systematic distractor (incorrect option) generation designed to match the plausibility of the correct answer~\cite{zang_are_2025}.

Most benchmarks assess a single musical modality: audio~\cite{weck_muchomusic_2024,zang_are_2025,koh_jamendo-qa_2025,lin_factual_2025}, visual notation (PDF/JPG score)~\cite{mundada_wildscore_2025,chen_musixqa_2025,yue_mmmu_2024,yue_mmmu-pro_2025}, or symbolic representation, either ABC notation~\cite{yuan_chatmusician_2024,zhao_abc-eval_2025} or multiple symbolic formats (ABC, Humdrum, MEI, MusicXML)~\cite{pond_teaching_2025}. A few recent benchmarks cover two modalities: SSMR-bench~\cite{wang_towards_2025} and MSU-bench~\cite{dai_musical_2025} jointly target symbolic and visual notation, while MuseBench~\cite{zhao_museagent-1_2026} covers audio, sheet images, and textual knowledge questions, though cross-modality performance is not comparable due to differing question types. To our knowledge, no benchmark evaluates MLLMs across more than two musical modalities.

Benchmark construction involves a trade-off between scale and quality. Human-curated benchmarks are small (e.g., 9 questions in~\citet{pond_teaching_2025}; 372 QAs in MusicTheoryBench~\cite{yuan_chatmusician_2024}), costly to produce, and may still be solvable without perception~\cite{weck_hummusqa_2026}. An alternative to manual annotation are automated approaches, such as converting caption datasets to QA via LLMs~\cite{weck_muchomusic_2024}, using predefined templates with manual~\cite{dai_musical_2025,zhao_museagent-1_2026} or programmatic ground truth~\cite{wang_towards_2025}, or generating questions from synthetic musical data~\cite{chen_musixqa_2025}, which yield larger benchmarks (e.g., MusiXQA with 130k QAs~\cite{chen_musixqa_2025}), but raise questions about validity of the QAs (when LLMs are involved in their creation) or transferability to real music (when synthetic musical data is used). Benchmark size is rarely discussed explicitly: human-curated sets are too small for stable estimates, while automatically generated sets are often too large to permit cost-effective evaluation of expensive models.

Finally, all existing benchmarks are static. Results may not transfer to music beyond the included examples, benchmark content can leak into future models' training data invalidating comparisons, and large benchmarks are resource-intensive to run --- a significant problem given how quickly new MLLMs are released. For a domain as diverse as music, a truly universal benchmark may be unattainable; even if one were constructed, new models would likely emerge faster than they could be evaluated.

\section{Methodology}
\label{sec:methodology}

\musicabench{} is organized around a predefined set of \metaq{}s (Tab.\ref{tab:meta-question-table}), each with an implemented function that extracts the ground truth and distractor options from a \texttt{MusicXML}. The resulting question-option pairs apply across symbolic files, sheet images, and audio, provided the dataset contains piece-aligned data in those formats, enabling cross-modal performance comparison.

Unlike SSMR-bench~\cite{wang_towards_2025}, which also programmatically extracts ground truth from symbolic notation (ABC notation), \musicabench{} does not produce a single static benchmark. Instead, it is a pipeline for benchmark instance generation over arbitrary user-provided data, and extends evaluation to the audio modality. 

\subsection{Question Template Design}

We ground our question templates in music theory, specifically in the foundational knowledge that marks the transition from intuitive listening to formalized structural awareness \cite{sloboda1985}. This analytical threshold, corresponding to the ``analysis'' level in Bloom's taxonomy \cite{bloom1956}, is central to major music pedagogy systems (Kodály \cite{Dobszay1972}, ABRSM, GCSE) and provides a shared vocabulary for musical communication. 

One author with formal musical training selected skills reflecting standard music learning evaluation structures (see \Cref{tab:meta-question-table}), then designed \metaq{}s satisfying four requirements: (i) answerable from all modalities (audio, image, symbolic); (ii) ground truth programmatically extractable from \texttt{MusicXML}; (iii) instantiable into multiple questions per piece; (iv) piece-level, i.e., answerable with the whole piece as context.

The skill categories also carry diagnostic value: e.g., low accuracy on audio pitch recognition makes successful audio-to-score transcription unlikely;
performance on the image modality may indicate OMR capability, etc.

\begin{table*}[t]
\centering
\small
\input{tab-meta-questions-v3}
\caption{Taxonomy of feature-recognition (perception) skills/categories and corresponding question examples (\textbf{bold} = variable values), with example answer options (final benchmark item contains 5 options always, see \cref{sec:evaluation-methodology}).}
\label{tab:meta-question-table}
\end{table*}

\subsection{Meta-Benchmark Generation Pipeline}
\label{sec:meta-benchmark-pipeline}

\begin{figure*}[]
    \centering
    \includegraphics[width=1\textwidth]{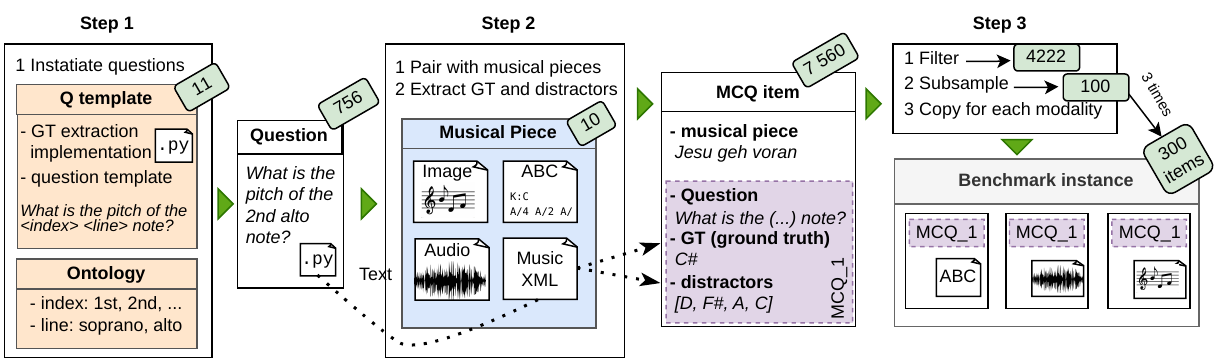}
    \caption{\musicabench{}: pipeline of constructing a benchmark instance. Inputs: (i) provided by us (orange): ontology, question templates with implementations of ground truth (GT) and distractor extraction methods; (ii) user-provided (blue): dataset. MCQ = multiple-choice question. Number of intermediate items for benchmark generation from ChoraleBricks dataset (see \Cref{sec:case-study-instance-for-chorale-bricks}) is shown in green boxes. For subsampling, desired target size $s=\qsPerSubcategory{20}$ is displayed, as used in \Cref{sec:model-comparison-on-choralebricks}.}
    \label{fig:pipeline-schema}
\end{figure*}

The pipeline (see \Cref{fig:pipeline-schema}) takes three inputs: the set of predefined \metaq{}s (see \Cref{tab:meta-question-table}), an ontology (a dictionary of possible values for evaluated musical elements\footnote{The ontology for tonal system datasets is automatically generated from \texttt{music21} library objects, mapping internal element names to the formal nomenclature used in questions and answer options.}), and most importantly, the musical data from which the benchmark is built. 
In \textbf{step 1}, \metaq{}s are instantiated into questions by replacing wildcards with all combinations of ontology values.
In \textbf{step 2}, a Cartesian product of questions and musical pieces is formed. For each (question, piece) pair, a \texttt{Python} extraction function receives the piece's \texttt{MusicXML} representation and the instantiating ontology values, and returns a ground truth answer and a pool of distractor options.\footnote{Distractors are drawn from the piece itself (e.g., pitches of subsequent soprano notes when querying the first soprano pitch); ontology-defined distractors serve as a fallback when too few options are available.}
In \textbf{step~3}, filtering removes invalid items (e.g., those querying a note index that does not exist in a given piece), subsampling balances the benchmark by skill/category and reduces the benchmark size (see \cref{sec:exps-10times-methodology-calibration-of-benchmark-size}), and finally, each (question, piece) pair is crossed with all modalities — symbolic (\texttt{ABC}), audio, and sheet image — so the same question is posed with the musical material in each modality, enabling direct cross-modal performance comparison.

Each benchmark item is a 4-tuple: (question, musical file, correct answer, 4 distractor options),\footnote{Distractors sampled randomly from the piece-specific distractor pool.} where musical files are unimodal — no item combines, e.g., audio and image simultaneously.

\subsection{Evaluation Methodology}\label{sec:evaluation-methodology}

We adopt a closed-question multiple-choice paradigm (reliability and reproducibility) while partially mitigating cue-reliance via a ``none of the other options is correct'' (NOTA) option.

All experiments use 5 options (1 true, 3 distractors + NOTA). NOTA is the correct answer in 20\% of questions (matching the $1/5$ uniform prior), where ground truth is replaced by an additional distractor. Options are shuffled randomly at generation time \cite{weck_muchomusic_2024,weck_hummusqa_2026}, as (M)LLMs are sensitive to option ordering \cite{lin_hearing_2026}. Responses are parsed to options (A–E) using regex matching and prompt formatting from MMMU-Pro \cite{yue_mmmu-pro_2025}.

\newcommand{\model}[1]{#1}

Models were selected for ease of use, performance, modality coverage, cost, and speed, and accessed primarily via the OpenRouter API
(except Qwen3 Omni 30B, run locally). The evaluated MLLMs are: Gemini 2.0 FL, 2.5 FL, 3.1 FL (Flash Lite; 3.1 in preview), Gemini 3.1 Pro (preview), and Qwen3 Omni 30B (Qwen3 Omni 30B A3B Thinking). For GPT and Mistral, which lack a single model covering all modalities, we use complementary pairs reported as unified systems prefixed ``agg-'': \model{agg-GPT-4o} (\texttt{gpt-4o-audio-preview} / \texttt{gpt-4o}), \model{agg-GPT-5} (\texttt{gpt-audio-mini} / \texttt{gpt-5-image-mini}), and \model{agg-Mistral} (\texttt{voxtral-small-24b-2507} / \texttt{mistral-small-3.2-24b-instruct}), where the first model handles audio and the second handles visual/symbolic input.

\newcommand{\fullBenchmarkSizeOneModality}[0]{4,222}
\newcommand{\fullBenchmarkSizeAllModalities}[0]{12,666}

\section{Dataset: ChoraleBricks}
\label{sec:case-study-instance-for-chorale-bricks}

We use the ChoraleBricks dataset \cite{balke_choralebricks_2025} as input for \musicabench{} for most of the experiments. It includes multitrack recordings of ten chorales, each with soprano, alto, tenor, and bass parts played by a variety of wind instruments, along with sheet music in MusicXML and PNG, ideal for \musicabench{}. For generating a \musicabench{} instance from ChoraleBricks, we use MusicXML for ground truth extraction, convert it to ABC notation\footnote{Python \texttt{abc\_xml\_converter} library.} for symbolic modality, and use reference mixes for audio, and sheet music PNGs for visual modalities.\footnote{both can be accessed at \url{https://audiolabs-erlangen.de/resources/MIR/2025-ChoraleBricks}} See \Cref{fig:pipeline-schema} for a pipeline schema, showing item counts in ChoraleBricks benchmark generation.

\section{Calibration of Benchmark Size}
\label{sec:exps-10times-methodology-calibration-of-benchmark-size}

The full ChoraleBricks-derived benchmark would contain
\fullBenchmarkSizeAllModalities{} items (\fullBenchmarkSizeOneModality{} $\times$ 3
modalities), making exhaustive evaluation computationally impractical.
Selecting an appropriate benchmark size requires balancing statistical power against
evaluation cost. Because MusICA-MetaBench targets real-world model selection rather
than fine-grained comparison, users need not detect arbitrarily small accuracy
differences—price, speed, and openness also inform choice. We therefore frame the
problem as: for a user-defined minimum relevant effect size $E$ (e.g., $E = 5\%$),
what is the smallest benchmark size that reliably detects a difference of $E$
percentage points?

\subsection{Calibration Methodology}

For each candidate size $s \in \{\qsPerSubcategory{1}, \qsPerSubcategory{5},
\qsPerSubcategory{10}, \qsPerSubcategory{20}, \qsPerSubcategory{50},
\qsPerSubcategory{100}\}$ (1, 5, 10, 20, 50, or 100 questions per category-modality combination), we
draw 10 random subsets without replacement using distinct random
seeds.\footnote{Subsets are balanced by categories, and such that each question appears in all modalities or none. Average pairwise overlap: 1.8\% for $s=\qsPerSubcategory{1}$, 5\% for $s=\qsPerSubcategory{20}$, 15.9\% for $s=\qsPerSubcategory{100}$.}
Evaluating several models on these subsets yields 10 accuracy measurements per model
per size. Normality verified by Shapiro-Wilk test \cite{shapiro_analysis_1965},
vast majority of cases passing.

To assess detectability, we shift the accuracy vector of the hypothesised superior
model so that the mean gap equals exactly $E$~\% while preserving the original
variance structure, then apply a paired $t$-test to the shifted values, and repeat this every ordered model pair across
$E \in \{1, 2, \dots, 20, 25, 30\}$~pp. For each $E$, we identify the smallest $s$
at which all pairwise tests are significant ($\alpha = 0.05$, Bonferroni-corrected).

\begin{table}[]
    \centering
    \input{tab-effect-size-vs-benchmark-size}
    \caption{Minimum detectable effect size (in \% of model accuracy
    difference) for each benchmark size $s$. 
    }
    \label{tab:effect-size-to-benchmark-size}
\end{table}

\subsection{Calibration Results}
\label{sec:size-calibration-results}

\begin{figure}
    \centering
    \includegraphics[width=1\linewidth]{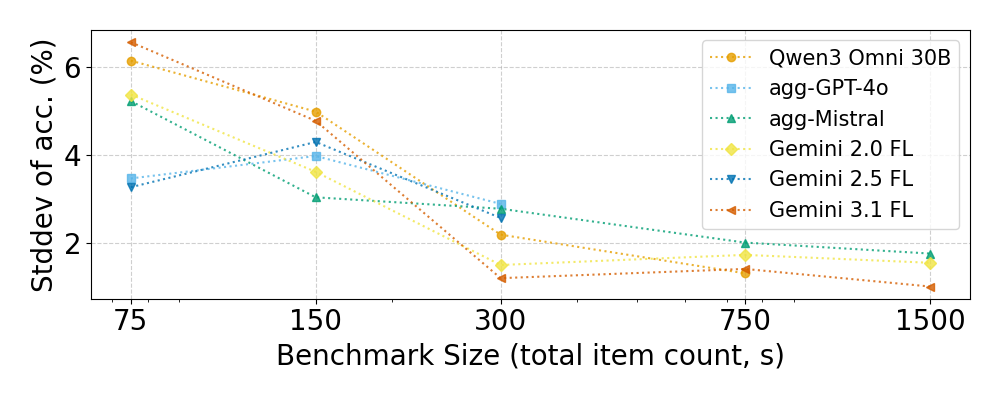}
    \caption{Standard deviation of model accuracy across 10 randomly subsampled
    benchmark instances for each benchmark size. Some models were excluded from larger
    sizes due to computational constraints. 
    X-axis log-scaled.}
    \label{fig:stddev-by-size-across-10-runs}
\end{figure}

We used 6 MLLMs for (see \Cref{fig:stddev-by-size-across-10-runs}
and \Cref{sec:evaluation-methodology}).
\Cref{tab:effect-size-to-benchmark-size} shows the minimum detectable
effect size per benchmark size. A size of $s = \qsPerSubcategory{20}$ suffices for a 6\% effect size. Validation on observed model differences confirms this: all accuracy gaps of at least 6\% are significant at $s = \qsPerSubcategory{20}$, and in practice even gaps of at least 3.5\% are.

\Cref{fig:stddev-by-size-across-10-runs} reports accuracy standard deviation across
10 subsampled instances per model-size combination. 
Standard deviation decreases as benchmark size grows, while mean
accuracy remains stable, supporting the hypothesis that a practically small benchmark
yields a reliable estimate of full-benchmark performance.

\section{Is Perception Required?}
\label{sec:normal-no_input-noise_input}

To validate our distractor generation, we test whether answering benchmark questions
requires perception of the musical content; that the benchmark does not suffer
from the ``not really listening'' problem \citep{zang_are_2025}, extended to ``not looking at the image'' and ``not reading the symbolic score'' in our multimodal setup). 
We compare MLLM performance against two ablation setups:

\begin{itemize}[noitemsep,nosep]
    \item \noinput{} (text-only): the musical file is omitted entirely; compatible with
    both MLLMs and text-only LLMs (cf.\ \citep{zang_are_2025}).
    \item \noiseinput{}: 
    Gaussian noise replaces audio/image, and a structurally valid but lexically nonsensical ABC score replaces the
    symbolic modality.
\end{itemize}

\begin{figure}
    \centering
    \includegraphics[width=0.9\linewidth]{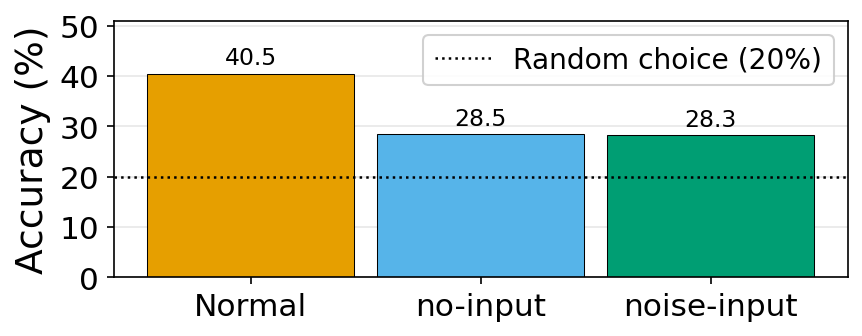}
    \caption{Robustness test results: is perception required? Normal setup vs. \noinput{} (no musical file is sent) and \noiseinput{} (Gaussian noise is sent instead of the musical file). Average accuracy over two MLLMs (Gemini 2.0 FL and Gemini 3.1 FL) on 10 independent instances of the benchmark of size $s=\qsPerSubcategory{20}$ reported.}
    \label{fig:normal-vs-textonly-vs-whitenoise}
\end{figure}

In \Cref{fig:normal-vs-textonly-vs-whitenoise} we observe a clear performance drop when switching from the normal setup to \noinput{} or \noiseinput{}. Both baseline setups are above random chance, however, which indicates 
we could do better in the choice of distractors: e.g., using the methodology from \citep{zang_are_2025}. %

\section{Benchmark Test-Drive}
\label{sec:test-drive}

We demonstrate \musicabench{} in practice on the ChoraleBricks and ChoraleSynth datasets.

\subsection{Model Comparison Results on ChoraleBricks}
\label{sec:model-comparison-on-choralebricks}

\begin{figure*}
    \centering
    \includegraphics[width=1\linewidth]{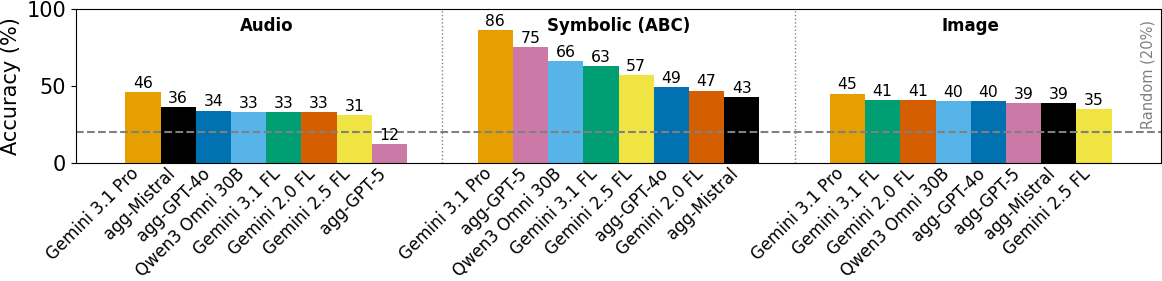}
    \caption{Per-modality results, ChoraleBricks, $s = \qsPerSubcategory{20}$,
    one benchmark instance.}
    \label{fig:results-bars-choralebricks-cross-modalities}
\end{figure*}

We set the minimum effect size of interest to 6~\% and consider three modalities:
audio recording, sheet image (PNG), and ABC notation. Based on
\Cref{sec:size-calibration-results}, this requires $s = \qsPerSubcategory{20}$
(300 questions total, 100 per modality). We generate one benchmark instance using
a fresh random seed and evaluate 8 MLLMs for overall (\Cref{tab:res-choralebricks-overall}), and per-modality accuracy (\Cref{fig:results-bars-choralebricks-cross-modalities}).\footnote{For per-category plot, see README.}

\begin{table}[]
    \centering
    \input{tab-s20_seed52_comparison}
    \caption{Results on Chorale-bricks: overall accuracy, runtime, and cost on ChoraleBricks,
    $s = \qsPerSubcategory{20}$, one benchmark instance.
    Qwen3 is run locally. Random baseline: 20.0 \%.}
    \label{tab:res-choralebricks-overall}
\end{table}

Gemini 3.1 Pro (released in preview on Feb 19th, 2026) is the clear winner.
Its margin of improvement suggests that music perception skills previously out of
reach for MLLMs \cite{DBLP:conf/ismir/MaLYBM25} may soon become attainable --- reinforcing
the need for flexible, actionable benchmarking as
we may be entering a period of time where \textit{some} MLLMs will become competitive for \textit{some} music perception applications, 
and decision-making between specialised
models and MLLMs may grow more complex.

Gemini 3.1 Pro is, however, astronomically expensive and slow compared to all other
models. If speed matters, Gemini 3.1 FL is the likely choice, though the benchmark
should be re-run at a larger size before concluding: the accuracy gap between
Gemini 3.1 FL and the next fast competitor, GPT-4o, is not safely significant
at $s = \qsPerSubcategory{20}$ (see \Cref{tab:effect-size-to-benchmark-size}).

Cross-modal results (\Cref{fig:results-bars-choralebricks-cross-modalities}) show
that most performance variation across models stems from the symbolic modality. In
audio, all models perform similarly except Gemini 3.1 Pro and GPT-5, the latter
failing to reach random-chance performance. Image modality scores are closely
clustered across models. Averaged across models,
$\text{accuracy}(\text{symbolic}) > \text{accuracy}(\text{image}) >
\text{accuracy}(\text{audio})$.

\subsection{Generalisation to Another Dataset: ChoralSynth}

To verify \musicabench{} generalises beyond ChoraleBricks, we apply it to
ChoralSynth \cite{narang_choralsynth_2023}, a synthetic multitrack choral dataset
(20 pieces). Ground truth and distractors are extracted from the provided MusicXML and converted to ABC notation
for the symbolic modality. We mix the audio tracks to a single recording per piece;\footnote{Valid
because the synthesized tracks are perfectly aligned.} render PDF score from
MusicXML via MuseScore and concatenate into a single PNG per piece.

Repeating the 10-run size calibration (2 models: Gemini 2.0 FL and Gemini 3.1 FL),
standard deviation decreases with benchmark size as on ChoraleBricks, but is
slightly higher at fixed sizes, which we attribute to greater item variance arising
from the larger number of distinct pieces in ChoralSynth.

On a single benchmark instance ($s = \qsPerSubcategory{20}$) with the same 8 MLLMs, the normal setup
achieves 28\% mean accuracy versus 17\% for \noinput{} and 20.1\% for \noiseinput{}
(both near the random baseline of 20\%), confirming perception is required.

\begin{table}[]
    \centering
    \input{tab-choral-synth-s20_seed52_comparison}
    \caption{Accuracy (\%) overall and per modality on ChoralSynth,
    $s = \qsPerSubcategory{20}$. Only the 3 best of 8 evaluated models are shown.
    Random baseline: 20\%.}
    \label{tab:results-choralsynth-overall}
\end{table}

Full results are in \Cref{tab:results-choralsynth-overall}. Gemini 3.1 Pro is again
the top model, but overall performance is substantially lower than on ChoraleBricks.
Performance drops are most pronounced in the audio and image modalities. For images,
the likely cause is that concatenated scores sometimes span multiple pages, making
them impractical to read. For audio, voice-specific labels (e.g.,
``Bassus~Ch1-F4'') may be impossible to identify from a mixed recording.

\subsection{Guidelines for Custom Datasets and Formats}

\musicabench{} supports any classical tonal repertoire consisting of one or more
monophonic voices (choral, strings, winds, etc.). 
For such datasets, the ontology mapping the system's chords, intervals, and rhythms, is generated automatically
from music21.

MusicXML is required for ground truth extraction. Other symbolic formats (MIDI, ABC
notation, LilyPond, Humdrum, etc.) may also work but have not been verified and may
require adjustments. Non-symbolic modalities must be piece-aligned with the MusicXML
source (same piece, same key, etc.). Modalities can vary in format and quality
(e.g., rendered, printed scan, or handwritten scan of a score) and can be compared
as submodalities, provided all are aligned to the same MusicXML file. Missing
modalities may be synthesised (e.g., rendering a score image or converting to ABC
notation). New question templates can be added, but require implementing the
corresponding ground truth extraction method.\footnote{For technical details on custom datasets, formats, and new templates, see corresponding README in Supplementary materials.}

\section{Conclusions and Future Work}
\label{sec:discussion}

\musicabench{} is a programmatic pipeline that generates QA benchmarks from user-provided musical data, targeting musical perception across modalities (audio, sheet images, symbolic formats) and scalable to optimal size. It offers an alternative to large static benchmarks, avoiding data leakage and generality constraints. We provide proof-of-concept evaluations on ChoraleBricks and ChoralSynth, 
noting that the new Gemini 3.1 Pro MLLM achieves a significant leap in musical capability on both. 
The pipeline is configurable and extensible to new datasets and formats. 
Total LLM experiment costs were 730 USD, dominated by leading-edge models.

\paragraph{Limitations.} Distractor difficulty may be slightly overestimated in the \noinput{} condition: the prompt instructs models to analyze a provided musical work, causing some to refuse when no file is present; a prompt such as ``try to guess the correct answer'' would be more appropriate.\footnote{The \noiseinput{} setup is unaffected.} Individual skill scores are not yet mapped to grading schemes such as ABRSM. Extending the pipeline to new datasets may require adjustment of ground-truth extraction methods, as validation was limited to ChoraleBricks and ChoralSynth.

\paragraph{Future work} includes harder distractor selection~\cite{zang_are_2025}, open-ended question formats, an extended \metaq{} set, musically multimodal questions requiring simultaneous perception of audio and notation, questions with multimodal answer options (e.g., multiple audio clips), and extension to polyphonic instruments (piano) and beyond tonal music.

As the musical capabilities of MLLMs grow, assessing what they can and cannot do in individual use-cases is becoming an important part of music processing system design. We believe the paradigm of \musicabench{} will be a practical way to obtain actionable insights on what ``AI'' really is (and isn't) capable of.

\section{Acknowledgments}

This work was supported by the project ``Human-centred AI for a Sustainable and Adaptive Society'' (reg. no.: CZ.02.01.01/00/23\_025/0008691), co-funded by the European Union, and partially by the SVV project number 260 821. 
The work described herein has been using services provided by the LINDAT/CLARIAH-CZ Research Infrastructure (https://lindat.cz), supported by the Ministry of Education, Youth and Sports of the Czech Republic (Project No. LM2023062).

\section{AI Usage Statement}
Large language models were used in three capacities in this work. First, as coding 
assistants during development of the \musicabench{} pipeline, including code 
drafting, debugging, and refactoring. Second, as writing assistants for drafting 
and rewriting portions of this manuscript. All AI-assisted outputs were critically 
reviewed, edited, and verified by the authors, who take full responsibility for 
the content of this paper. Third, and most centrally to this work, 
(M)LLMs served as the \textit{subjects of evaluation}: the models assessed by 
\musicabench{} are themselves large language models, evaluated on their 
music perception capabilities as described in the experimental sections.
The pipeline itself (both benchmark generation and response evaluation) is programmatic and LLM-independent.

\section{Ethics Statement}

\paragraph{Data and copyright.}
\musicabench{} is demonstrated on the ChoraleBricks and ChoralSynth datasets, 
both of which are publicly available research datasets distributed under open licences. A key motivation of our framework is precisely to avoid 
the copyright concerns inherent in large static benchmarks: by enabling researchers 
to generate benchmarks from their own data, \musicabench{} shifts responsibility 
for data licensing to the end user, who evaluates models on material they are 
already authorised to use.

\paragraph{Bias and generalisability.}
Our question templates are grounded in Western tonal music theory and pedagogy 
(e.g., Kod\'{a}ly, ABRSM, GCSE frameworks). The framework therefore reflects a 
culturally specific analytical tradition and may not generalise to non-Western 
musical systems, microtonal music, or oral traditions. Users should be aware of 
this scope when drawing conclusions about model capabilities.

\paragraph{Intended and unintended use.}
\musicabench{} is intended as a diagnostic tool for MIR researchers assessing 
MLLM readiness for specific tasks, not as a definitive ranking of model quality. 
Misuse of benchmark results to make broad claims about model superiority, or to 
guide high-stakes deployment decisions without additional validation, is 
discouraged.

\paragraph{Environmental cost.}
Total API costs for the reported experiments were approximately 730~USD, dominated 
by frontier models. We encourage users to apply the benchmark-size calibration 
procedure described in this paper to minimise unnecessary inference compute.

\paragraph{Human subjects.}
No human subjects were involved in this research. No personally identifiable 
information was collected or processed.

\bibliography{references,k_references}

\end{document}

%% file: tab-meta-questions-v3.tex
\newlength{\tablelinespace}
\setlength{\tablelinespace}{0.2em}

\begin{tabularx}{\textwidth}{@{} p{1.4cm} @{\hspace{0.35em}} X p{3.7cm} @{}}
\toprule
\textbf{Skill} & \textbf{Question Example} & \textbf{Options Examples} \\
\midrule

\multirow{2}{=}{Pitch} &
\textbf{Q1.} What is the pitch name of the \textbf{4th} \textbf{soprano} note in the provided excerpt? &
C\#, G, D, B$\flat$ \\
\addlinespace[\tablelinespace]

&
\textbf{Q2.} How many times is the note \textbf{F} present in the \textbf{alto} line? &
12, 3, 5, 8 \\
\midrule

\multirow{2}{=}{Interval} &
\textbf{Q3.} What is the \textbf{2nd} interval in the \textbf{bass} line? &
Minor Third, Major Second \\
\addlinespace[\tablelinespace]

&
\textbf{Q4.} How many times is the interval \textbf{Perfect Fourth} present in the \textbf{tenor} line? &
2, 0, 5, 3 \\
\midrule

\multirow{2}{=}{Rhythmic} &
\textbf{Q5.} What is the rhythmic notation of the \textbf{10th} \textbf{soprano} note (or rest)? &
whole, eighth, sixteenth \\
\addlinespace[\tablelinespace]

&
\textbf{Q6.} What is the total count of \textbf{whole}-note units in the \textbf{soprano} line? &
4, 0, 7, 2 \\
\midrule

\multirow{2}{=}{Temporal Proportion} &
\textbf{Q7.} What is the temporal relationship (e.g., 2:1, 3:1) between the \textbf{3rd} note (or rest) and the next one in the \textbf{alto} line? &
1:2, 2:3, 1:1, 3:1 \\
\addlinespace[\tablelinespace]

&
\textbf{Q8.} How many rhythmic figures with a strict \textbf{1:2} proportionality are there in the \textbf{bass} line? &
1, 4, 7, 0 \\
\midrule

\multirow{3}{=}{Harmonic} &
\textbf{Q9.} What is the \textbf{tonic} harmony of the primary key of this piece? &
G Major, C Minor, D Major \\
\addlinespace[\tablelinespace]

&
\textbf{Q10.} What is the chord at the \textbf{5th} position considering all the voices? &
Minor Sixth, Diminshed triad \\
\addlinespace[\tablelinespace]

&
\textbf{Q11.} What is the chord progression between the \textbf{8th} chord and the next one? &
V~--~I, IV~--~V, ii~--~V, vi~--~IV \\

\bottomrule
\end{tabularx}

%% file: tab-effect-size-vs-benchmark-size.tex
\begin{tabular}{lllllll}
  \toprule
  benchmark size & \qsPerSubcategory{1} & \qsPerSubcategory{5} & \qsPerSubcategory{10} & \qsPerSubcategory{20} & \qsPerSubcategory{50} & \qsPerSubcategory{100} \\
  min effect (pp) & 30 & 12 & 8 & 6 & 3 & 2 \\
  \bottomrule
\end{tabular}

%% file: tab-s20_seed52_comparison.tex
\begin{tabular}{llrr}
\toprule
Model & Acc. (\%) & Time & Price (\$) \\
\midrule
Gemini 3.1 Pro & 59.0 & 12 h & 58.4 \\
Qwen3 Omni 30B & 46.3 & 2 h & N/A \\
Gemini 3.1 FL & 45.7 & 18 m & 0.3 \\
agg-GPT-5 & 42.0 & 91 m & 1.3 \\
agg-GPT-4o & 41.0 & 21 m & 2.6 \\
Gemini 2.5 FL & 41.0 & 90 m & 0.6 \\
Gemini 2.0 FL & 40.3 & 16 m & 0.04 \\
agg-Mistral & 39.3 & 89 m & 0.5 \\
\midrule
Random Baseline & 20.0 & N/A & N/A \\
\bottomrule
\end{tabular}

%% file: tab-choral-synth-s20_seed52_comparison.tex
\begin{tabular}{lllll}
\toprule
Model & All & Audio & ABC & Image \\
\midrule
Gemini 3.1 Pro & 47.0 & 31 & 75 & 35 \\
Gemini 3.1 FL & 33.3 & 23 & 49 & 28 \\
Gemini 2.5 FL & 32.0 & 22 & 47 & 27 \\
\bottomrule
\end{tabular}